# Observations on spectral deformations of $^{103m}$Rh excited by bremsstrahlung

Y Cheng[1], B Xia[1], Q-X Jin[1], and C P Chen[2]

[1] Department of engineering Physics, Tsinghua University, 100084, Beijing, China
[2] Department of Physics, Peking University, 100871, Beijing, China

E-mail: yao@tsinghua.edu.cn

**Abstract.** Spectral deformation of K$\alpha$, K$\beta$ and $\gamma$ emissions from the nuclear state $^{103m}$Rh excited by bremsstrahlung are investigated. Nonlinear increase for excitation number density of $^{103m}$Rh with radiation exposure is observed. The spectral profiles are broadened, attributable to a triplet splitting. Interesting time-evolution behaviors of the spectral deformations are obtained.

## 1. Introduction

In this paper, we report the time-resolved observations at room temperature on the spectral deformation of the K lines and $\gamma$ ray from the nuclear state $^{103m}$Rh excited by bremsstrahlung. The excited $^{103m}$Rh is a long-lived nuclear state of 39.76 keV with a half life of 56 minutes. Some of the anomalous phenomena on the $^{103m}$Rh excited by bremsstrahlung have been reported in our previous papers [1, 2]. In these works, the decay speeds of the K lines and the $\gamma$ ray are found different in the regime of relatively low bremsstrahlung excitation, in which regime, the excitation number density (inversion density) of $^{103m}$Rh increases linearly with the exposure of bremsstrahlung irradiation. In this report, we discover a trend of nonlinear increase of the inversion density beyond $3\times10^{11}$ cm$^{-3}$. This nonlinear increase is not an isolated observation but accompanied with an irreversible enhancement of the spectral deformations observed for the time-resolved behavior of all of the three characteristic emissions. In the linear regime, the time-resolved analysis of spectral deformations reveals that a broadening on the profiles of the characteristic emissions appears, attributable to a triplet splitting, and that the splitting is stationary with only a weak dependence on the inversion density. However, this stationary splitting significantly opens up for the spectral profiles in the nonlinear regime. The splitting magnitude remains at the same value corresponding to the nonlinear regime even when the emission luminosity relaxes down, corresponding to the level in the linear regime. On the other words, the splitting magnitude does not decrease reversibly down to the level corresponding to the linear regime during the period of three-hour measurement. The fast atomic emissions of x-rays in femto-second and the slow nuclear emission of $\gamma$ ray in kilo-second exhibit the same splitting behavior. Since the fast atomic process is dictated by the slow nuclear process in the experiment, the excitation by the bremsstrahlung on the atomic level is expected to relax away even before the measurement begins without yielding a significant contribution. It indicates that the broadening of the K$\alpha$, K$\beta$, and $\gamma$ emissions is of the same origin from the nuclear process. An atomic process, such as the self-absorption, is unlikely to give rise to any appreciable effect on the observed anomaly.

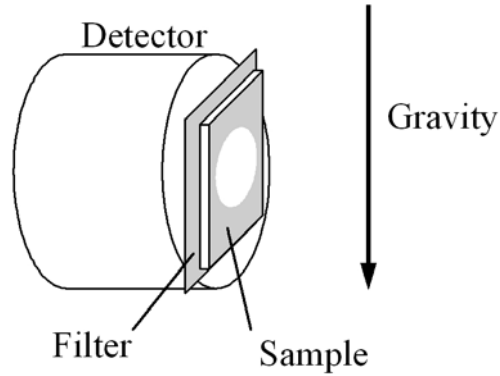

Figure 1. Simple diagram for configuration of measurement. The central white spot on the sample stands for the irradiation location. Its intensity distribution is Gaussian with a FWHM of 8 mm. The filter is a cooper sheet of 35 μm in thickness. The relative orientation of the set-up is indicated by the arrow for the gravity.

## 2. Experimental arrangements and measurements

The sample is made of polycrystalline rhodium with 99.9% purity (Goodfellow Rh00300). The natural abundance of $^{103}$Rh is 100% such that almost all of the nuclei in crystal are identical. It is in square shape with dimensions of 25×25×1 mm$^3$. To generate the excited state of $^{103m}$Rh, the sample is irradiated by the bremsstrahlung from a 6-MeV linac, as reported in Ref. [1,2]. The detector is a low-energy high purity germanium (HPGe, CANBERRA GL0510P) detector with 500 mm$^2$ active area directly linked with an optic feedback pre-amplifier (CANBERRA 2008 BSL) and is covered by a beryllium window. The data acquisition system consists of a linear amplifier (CANBERRA 2025) and a multi channel analyzer (MCA, CANBERRA Multiport II). The detector is horizontally leveled, as shown by the simple diagram in figure 1, and oriented along the north-south direction, roughly parallel to the earth magnetic field. The bremsstrahlung is generated by electron macro pulses of 4.5 μs with four different repetition frequencies, viz, 105, 210, 260, and 295 Hz. Its intensity, which is monitored by a dosemeter, is shown to linearly depend on the repetition frequency of the electron-beam for 105, 210, and 260 Hz. The bremsstrahlung intensity further increases from this linear proportionality with the frequency of 295 Hz by additionally increasing the beam energy. The irradiation spot is Gaussian with a full width half maximum (FWHM) of 8 mm in diameter at the sample center. The duration for the sample irradiation is 2 hours immediately followed by the time-resolved measurement in a period of 3 hours. In a typical measurement, 180 sub-spectra with a successive sequence in the time domain are obtained. Each sub-spectrum is taken with the duration of 1 min for data-taking. A full spectrum is thus obtained by adding up all of the sub-spectra which carry the time-resolved information. In order to suppress the Kα-Kα pile-up of 40.4 keV, which would otherwise appear at the right shoulder of the 39.76-keV γ peak, we insert a copper sheet of 35 μm between the rhodium sample and the HPGe detector. The Cu filter reduces the Kα luminosity by more than three times and thus suppresses the pile-up at 40.4 keV. To estimate the number of excited nuclear states in the sample, the following factors were considered: (i) the active area of the detector head gives the collection solid angle of about 1.2π sr; (ii) the transmission of rhodium Kα x-rays out of the sample were estimated as about 3.7% by using the absorption coefficient data from Ref. [3,4];

and (iii) the internal conversion rate of Rh Kα lines is 6.2%, according to Ref. [5]. Thus, by considering the Kα initial count rate of $5\times10^3$ count-per-second (cps) and the decay time constant of 4857 s, the total number of $^{103m}$Rh in the sample produced by the bremsstrahlung can be estimated as about $4\times10^{10}$. We assume the radial distribution of the irradiation in the sample is Gaussian, so the number of $^{103m}$Rh within the FWHM irradiation spot is estimated as $2\times10^{10}$. The corresponding inversion density is then estimated as about $4\times10^{11}$ cm$^{-3}$, which produces the initial count rate of $5\times10^3$ cps for the Kα luminosity. The energy spectra are taken with the channel width of 29.4 eV. However, we scale the normalized count by the factor of 25.7/29.4 for each channel in accordance with the normalized spectra of the future works.

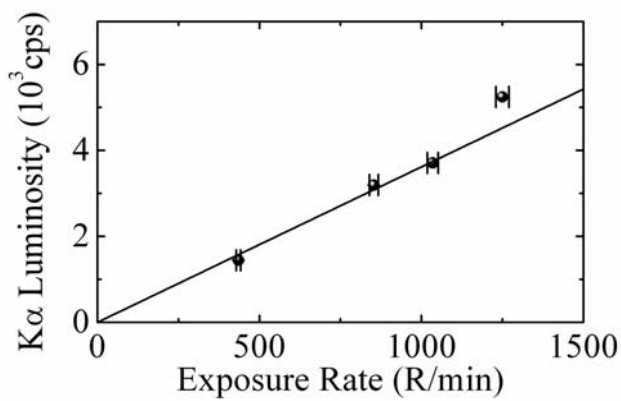

Figure 2. Kα luminosity obtained from the first minute measurement as a function of the recorded value of the exposure rate in Röntgen per minute (R/min). The repetition frequencies of the electron beam to generate the specified bremsstrahlung exposure rate are 105, 210, 260, and 295 Hz. The solid line is for the linear fit from the three data points below 1200 R/min.

In the experiments, the bremsstrahlung intensity is proportional to the exposure rate measured by the dosemeter (PTW UNIDOS) with a Farmer ionization chamber (PTW TW30013) located 1 meter behind the sample. The upper limit of the repetition frequency is 300 Hz. In order to obtain higher intensity of bremsstrahlung, the electron beam energy is increased by 2.5%, ~6.15 MeV, in the case of 295 Hz. The corresponding increase in the endpoint energy of the bremsstrahlung is insignificant to produce the photo-neutrons in the linac target consisting of a 1.5-mm W sheet and a 1-mm Au sheet laminated together. The unwanted photo-neutron would create the contamination of $^{104m}$Rh with the half life of 4.34 min, which has been observed to yield several cps, if the beam energy exceeds 6.2 MeV. All of the measurements reported in this work are therefore free from the contamination of $^{104m}$Rh. The measured luminosity of the initial Rh Kα count rate, which is determined from the first minute sub-spectrum, is expected to be proportional to the accumulated exposure from the two-hour irradiation. Figure 2 illustrates the measured luminosity for the initial Kα count rate as a function of the exposure rate. The luminosity increases linearly with the exposure rate below 1200R/min. However, a sign of nonlinear increase appears with the measurement by the repetition rate of 295 Hz above 1200 R/min. Further increase in the bremsstrahlung intensity by tuning the macro-pulse duration of the electron beam from 4.5 to 5.0 μs has been performed experimentally by a different

HPGe detector to confirm the nonlinear increase. The data are presented in figure S2 in the supporting online material. The corresponding detailed analysis will be reported in the future. The background count rate has been studied with the passive lead shielding of 10 cm in thickness. It is very low, about 0.1 cps, in the energy range of interest (0~40 keV). The major background peaks are from the K lines of the lead shielding. This count rate is far below that of our measurement by four orders of magnitude. On the other hand, the maximum count rate of the system is limited by the MCA, which is estimated as about $5\times10^4$ cps. The maximal count rate, $6.5\times10^3$ cps, at the beginning of our measurements is far below this limit with a dead time of 14%.

## 3. Calibrations and data analysis

*3.1 Calibrations of energy resolution and broadening of characteristic emissions*
In the first step, we compare the profiles of the K x-ray peaks emitted from $^{103m}$Rh excited by bremsstrahlung with those from the radioactive source of $^{109}$Cd. The $^{109}$Cd nucleus becomes $^{109m}$Ag by electron capture. The nuclear transition of $^{109m}$Ag($^{109}$Cd) is a multipolar E3 transition, which is of the same type as that of $^{103m}$Rh. The characteristic emissions of both $^{109m}$Ag($^{109}$Cd) and $^{103m}$Rh are K lines and γ, but with different energies and internal conversion ratio. The Kα lines at 22 keV and the Kβ lines at 25 keV from $^{109m}$Ag($^{109}$Cd) are slightly broadened on the high-energy side by the hypersatellites [6,7]. The intensity fraction of hypersatellites induced by the multiple ionizations of two K holes is on the order of $10^{-4}$. Our coincidence measurement also reproduces this value. On the other hand, the satellite lines from $^{103m}$Rh have never been reported in the literature. In our experiment, we have observed a broadening of the Rh K lines not only on the high-energy side but also on the low-energy side. This is a rather different feature from the Ag K broadening attributed to the hypersatellites.

The FWHM energy resolution of the HPGe detector has been calibrated by a series of radioactive sources elaborated in the next paragraph. The FWHM energy resolution is 385 eV for the Rh Kβ lines at 22.7 KeV, which is about 3% less than 396 eV for the Ag Kβ lines at 24.9 KeV. Moreover, the separation between Ag Kβ$_1$ and Kβ$_3$, ~31 eV, within the Kβ peak is larger than that of Rh, ~25 eV. According to these two facts mentioned above, the Ag Kβ peak consisting of Kβ$_1$ and Kβ$_3$ lines at 24.9 keV is expected to be broader than the corresponding Rh Kβ peak, also consisting of Kβ$_1$ and Kβ$_3$ lines, at 22.7 keV. In order to compare the magnitude of broadening between these two Kβ peaks in our experiments, we have normalized the peaks by the corresponding total count, and then shift the Ag Kβ$_3$ line to coincide with the Rh Kβ$_3$ line by linear interpolation, as shown in figure 3(a). In the figure, the Rh Kβ lines, which are buried within the same single peak due to the intrinsic energy resolution of the detector, are represented by the vertical solid lines and the Ag Kβ lines are represented by the vertical dotted lines. The ratio of intensity for the Rh Kβ peak over the Ag Kβ peak is then plotted in figure 3(b). The dashed curve in figure 3(b) shows the theoretically calculated value of this ratio. It takes into account the internal emission intensity, the self-absorption coefficient of the X-ray in different materials and the geometry of the samples for both of $^{103m}$Rh and $^{109}$Cd. Experimentally, this ratio is clearly enhanced as shown on the left hand side (lower energy side) of the spectrum. It indicates that the spectral profile of the Rh Kβ lines is broadened. The right shoulder, which is about 8% larger than 1.00 in figure 3(b), is caused by the stronger penetration of the Rh Kβ$_2$ photons in the Rh sample of 1 mm in thickness, which is thicker than that of the $^{109}$Cd source.

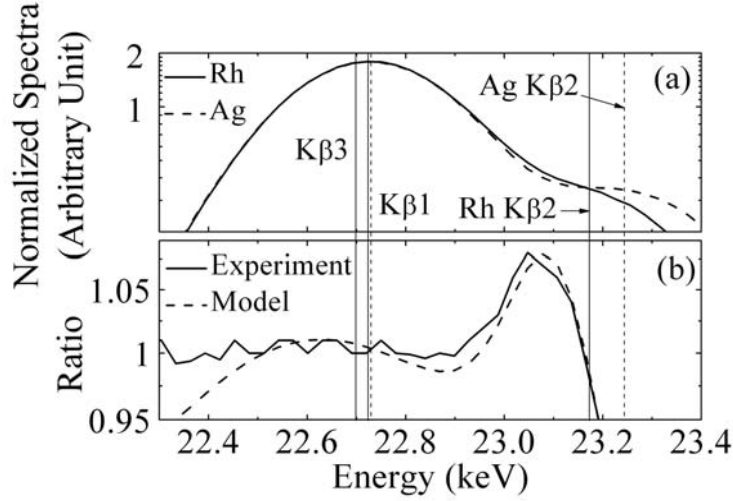

Figure 3. (a) Detected profiles of the Kβ lines for $^{103m}$Rh (vertical solid lines) and $^{109m}$Ag($^{109}$Cd) (vertical dotted lines). The Ag Kβ$_3$ line is shifted down to coincide with the Rh Kβ$_3$ line. The spectral profile of Rh (solid curve) is obtained by the bremsstrahlung irradiation with the repetition frequency of 210 Hz without inserting the copper filter and the profile of Ag (dashed curve) is obtained by the decay of radioactive $^{109}$Cd source. (b) The ratio of the Rh Kβ lines over the Ag Kβ lines. The solid curve stands for the experimental result and the dashed one is calculated according to the model.

In order to further investigate the broadening of Rh K peaks and γ ray, the normal profiles for an energy peak detected by the HPGe detector has been calibrated by the model provided by the vendor. A single energy peak is described by a Gaussian function with an additional tail in the low-energy side. The key parameter to be calibrated is the FWHM energy resolution, $\delta E_{FWHM}$, for the detected characteristic peaks at different energy. We select the following K peaks for the calibration, including Mo Kα (17.4 keV), Ag Kα (22.1 keV), and Ag Kβ (24.9 keV). The Mo Kα lines are produced by inserting a 25-μm Mo sheet between the excited Rh sample and the detector. The Ag K peaks come from the radioactive $^{109}$Cd source. Two parameters, $a$ and $b$, for the straight line $\delta E_{FWHM}(E) = a + b\sqrt{E}$ going through the three points of $\delta E_{FWHM}$ mentioned above, as listed in Table 1, are determined as $a = 280$ eV and $b = 0.698$ eV$^{1/2}$ by the fitting analysis, where $E$ is the energy of the calibration peaks. The value of $\delta E_{FWHM}$ for the normal profiles of Rh K peaks are then calculated by the formula, $a + b\sqrt{E}$. The normal profiles of Rh K peaks are also obtained experimentally by inserting an 100-μm Rh sheet between a radioactive source of $^{195m}$Pt and the detector. The Rh K fluorescence, which is generated by the $^{195m}$Pt γ ray at 30.9 keV rather than by the bremsstrahlung irradiation, is not broadened. It is noted that the normal profiles for the Rh K peaks excited by the $^{195m}$Pt γ ray are recorded using the same detector but with a different MCA, which has a different channel width and different resolution from the MCA for the rest of the measurements. The comparison for consistency between the data sets taken by the two different MCA is carried out by the same $^{109}$Cd source. Similarly, $\delta E_{FWHM}$ for the γ ray of $^{103m}$Rh at 39.76keV is also obtained by the same calibration procedure against the γ peaks of $^{241}$Am (26.34 and 59.5 keV) and $^{109}$Cd (88.03 keV). The individual Rh K line in a K peak is not resolvable by the HPGe detector. Therefore, the ratio between the indistinguishable K lines is an important parameter to model the normal profiles. To estimate the ratio, the intrinsic emission intensity of each K lines is adapted from the table [5], and then the self-absorption [3,4] of the Rh sample is taken into account. The self-absorption coefficients for these K lines are further confirmed by inserting external filters to observe the corresponding attenuation effect. The estimated ratios are listed in Table 2. The deformations of the spectral profiles are then analyzed by the normal profiles calculated above.

**Table 1.** Values of FWHM, $\delta E_{FWHM}$, for the normal profiles of K peaks from Mo and Ag, and for the Rh characteristic emissions. The FWHM for the K peaks of Mo and Ag is determined by the fitting analysis using a function for the normal profiles provided by the vendor. The FWHM for the Rh characteristic emissions is obtained by calculation using the calibration formula mentioned in the text.

|  | Mo Kα | Rh Kα | Rh Kβ | Ag Kα | Ag Kβ | Rh γ |
|---|---|---|---|---|---|---|
| Peak Energy (keV) | 17.4 | 20.2 | 22.7 | 22.1 | 24.9 | 39.76 |
| $\delta E_{FWHM}$ (eV) | 375±2 | 380±3 | 385±5 | 385±0.7 | 396±0.5 | 418±3 |

**Table 2.** Ratios between the individual Rh K lines for the normal profiles.

|  | Kα2/ Kα1 | Kβ3/ Kβ1 | Kβ2/Kβ1 |
|---|---|---|---|
| Estimated ratio | 0.488 | 0.513 | 0.264 |

*3.2 Spectral deformations and splitting models*

The normalized spectral deformations, which reveal the deviation of the measured spectral profiles from the calculated normal profiles, is formulated as,

$$\Delta S_i(E, t_m) = S_i(E, t_m) - \bar{S}_i(E), \quad (1)$$

in which $i$ stands for Kα, Kβ, and γ, $E$ is the energy of the characteristic emissions, $S_i(E, t_m)$ is for the sub-spectrum taken at the time sequence, $t_m$, with the data-taking time of 1 minute, and $\bar{S}_i(E)$ is for the normal profile with the calibrated FWHM listed in Table 1. Both the measured spectra $S_i(E, t_m)$ and the normal profile $\bar{S}_i(E)$ are normalized, respectively, by $\int \bar{S}_i(E) dE = 1$ and $\int \sum_m S_i(E, t_m) dE = \sum_m e^{-t_m/\tau}$, where $\tau = 4857$ s, is the decay time constant of $^{103m}$Rh.

Figure 4(a) shows the spectral deformations analyzed by (1) for the accumulated data taken within the entire period of measurement, i.e. for the full spectrum by the summation over all of the 180 sub-spectra in 3 hours. The sample is excited by the bremsstrahlung irradiation with the repetition frequency of 105 Hz, while the data is taken by inserting the Cu filter. In order to avoid any possible artifact arising from the calculated normal profile term in (1) by the calibration procedure for the energy resolution mentioned above, a differential analysis is also applied to validate the profile deformations. We replace the normal profile term, which is the second term in (1), by the measured spectral profiles of 105 Hz to obtain the differential mapping for the data taken by the repetition frequencies of 210 Hz and 260 Hz with Cu filter, as shown in figures 4(b) and (c), respectively. If the deformation profiles in figure 4(a) were attributed to any miscalculation of the normal profiles, similar deformation patterns would not have shown up in the differential mapping with both of the terms in (1) being the normalized spectra of experimental data. On the other hand, if the deformation profiles in figure 4(a) were attributed to the self-absorption of the Rh sample or the Cu attenuation, then the deformations in figures 4(b) and (c) would not have appeared. Since the self-absorption effect depends linearly on the x-ray intensity, any possible deformations caused by this effect will be normalized out. The *normalized* profiles, shown in figures 4(a), (b) and (c), are similar. This indicates that the spectral deformation is not an artifact.

For the spectral deformation of K lines, asymmetrical splitting appears, whereas the γ splitting is quite symmetrical. Nonetheless, it is difficult to positively indentify the symmetry for all of the γ splitting, as shown in figures 4(a), (b), and (c), owing to the fluctuation arising from the low total count of γ. The asymmetrical part of the K splitting is probably attributable to the satellite lines of multiple ionizations. However, the total deformation counts summing over the channels within the peaks, as shown in figure 4, are by two orders of magnitudes larger than the reported hypersatellite contribution from $^{109m}$Ag [6,7].

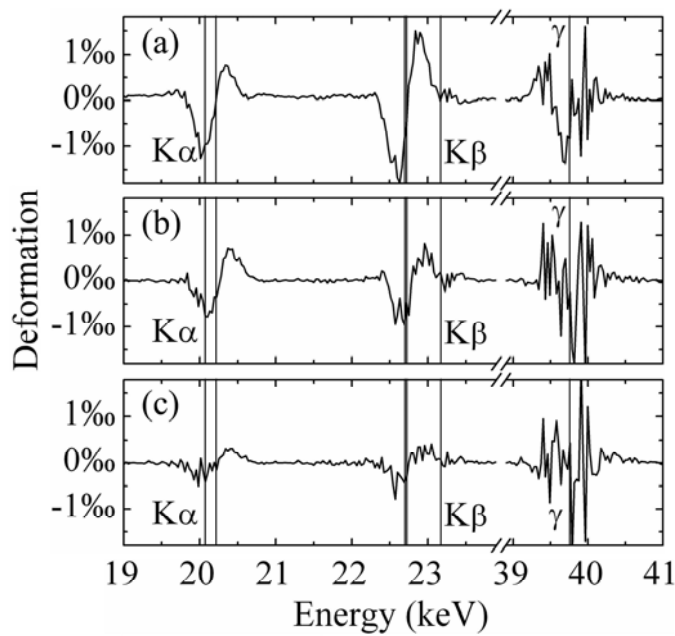

Figure 4. Normalized spectral deformations for the full spectrum accumulated in 3 hours for Kα, Kβ and γ peaks with the copper filter. The two vertical lines at the position of Kα are for Kα$_1$ and Kα$_2$. At the position of Kβ, the two vertical lines almost coincided are for Kβ$_1$ and Kβ$_3$, and the third one is for Kβ$_2$. The location of γ at 39.76 keV is also indicated by the vertical line. (a) The spectral deformations for 105 Hz obtained according to (1). (b) The differential mapping between the spectral profiles of 210 and 105 Hz. (c) The differential mapping between the spectral profiles of 260 and 105 Hz.

In order to further investigate the anomalous broadening of the characteristic emissions as revealed in the preceding subsection, the broadened Rh peaks with the energy splitting by the spectral deformation analysis above are analyzed by a model of triplet splitting. By this model, the three peaks of Kα, Kβ and γ in the full spectrum accumulated in 3 hours along with the corresponding triplet splitting curves are assumed with the normal profiles used in (1), formulated as,

$$S_i(E,t_m) = A_{c,i}(t_m)\overline{S}_i(E) + A_{l,i}(t_m)\overline{S}_i(E - \Delta E) + A_{r,i}(t_m)\overline{S}_i(E + \Delta E) \quad (2)$$

in which the subscript $i$ stands for Kα, Kβ or γ, the subscripts c, l, and r for the central, the left and the right peaks of the triplet splitting, respectively. There are 4 fitting parameters for the analysis, including 3 for the amplitudes represented by $A_{c,i}$, $A_{l,i}$, and $A_{r,i}$, and the magnitude of energy separation, $\Delta E$, for the splitting. With the constraint, $A_{c,i} + A_{l,i} + A_{r,i} = 1$, only 3 free fitting parameters are left. In the extreme limit with $A_{c,i} = 1$, $A_{l,i} = A_{r,i} = 0$, the detected characteristic peak reduces to a single peak exhibiting normal profile without any broadening effect. In the other extreme, $A_{c,i} = 0$, and $A_{l,I}$, $A_{r,i} > 0$, equation (2) describes a doublet splitting with the energy separation of 2×$\Delta E$ between the two peaks. For the intermediate state of triplet splitting, $0 < A_{c,i} < 1$, the magnitude of the splitting energy, $\Delta E$, is found to depend on the amplitudes, $A_{c,i}$, $A_{l,i}$, and $A_{r,i}$. The larger the magnitude of $\Delta E$, the larger the value of $A_{c,i}$. The optimum values for these parameters are then determined by the fitting analysis.

With the FWHM of the detector, ~ 400 eV, as listed in Table 1, it is impossible to resolve the spectral deformation with a splitting energy << 400 eV directly by the experimental measurement. By the data analysis, however, it is revealed that the spectral deformation for the broadening of the characteristic emission is describable by the triplet splitting state with $0 < A_{c,i} < 1$, and a splitting energy, $\Delta E$ << 400 eV. The splitting energy becomes larger for the measurements in the nonlinear regime. With $\Delta E$ > 400 eV, the triplet splitting is clearly observed as shown in figure S4 of the supporting online material. On the contrary, the triplet splitting with repetition frequency of 295 Hz is not resolvable, as shown in figure S5 of the supporting online material. Detail analysis in the nonlinear regime will be reported elsewhere. However, one of the important results from the analysis is that the amplitude is $A_{c,i}$ ~ 80% for the central peak of triplet. Though it is difficult to fully resolve the splitting parameters for the data in this work, we adopt two schemes for the analysis. One is to fix the central amplitudes as $A_{c,i}$ = 80% (T model). This yields a consistent picture with those of measurements in the nonlinear regime reported later. The other is to take the extreme condition with $A_{c,i}$ = 0, which is termed as the doublet splitting without the central peak (D model). The splitting energy is minimum for the case of doublet splitting. The results of analysis by these two schemes are listed in Table 3. The γ splitting energies for the three measurements in the linear regime are within the range from 40 to 60 eV, for the triplet model. An interesting enhancement in the splitting energy is observed for the spectral deformations by the excitation of 295 Hz, which is in the nonlinear regime as discussed in figure 2, in comparison with the other three data points in the linear regime. More data taken by a different HPGe detector in the nonlinear regime further confirmed this trend of increment, as presented in figure S3 of the supporting online material.

*3.3 Time-resolved spectral deformations*
The plots on the top of figures 5 and 6 depict the spectral deformations of the full spectra analyzed by (1) and (2) for 210 and 295 Hz without inserting the Cu filter, respectively. The corresponding fitting curves for T and D models are also presented in the figures. In both of the top plots, the curves for the spectral deformation of full spectrum analyzed from the spectra of measurements, with the T model in blue color and the D model in red color. Both of the calculated curves do not show much difference from each other. It indicates that the broadenings for these emissions are insensitive of whether it is analyzed by T or D models. This is attributed to $\Delta E$ << 400 eV. For the data in the nonlinear regime as shown in Figure S4, however, it is clear that triplet splitting model is more appropriate. The time-resolved spectral deformations of the 180 sub-spectra with the triplet splitting are shown in the bottom parts of these two figures. For the time-resolved analysis, the spectra with the data count per minute taken at the time, $t_m$, are analyzed. They demonstrate that the deformations are stationary for the data taken both in the linear and nonlinear regimes. Usually, most of the mistakes made in measurements or in the subsequent data analysis are manifested in the time-dependent or the time-resolved γ spectroscopy. This is not the case in our experiment, however. In addition, enhancement in the magnitude of the deformations is observed for the spectrum in the nonlinear regime by 295 Hz, shown in figure 6, in comparison with that in the linear regime shown in figure 5. In figure 6, asymmetrical splittings of the K peaks and the apparently symmetrical splitting of the γ peak are clearly revealed. The symmetrical splitting of γ ray provides a piece of strong evidence against the possibility of self-absorption of the single photon energy. Strikingly, this significant enhancement remains stationary during the three-hour measurement, even when the inversion density of $^{103m}$Rh decays to the level much lower than that in the linear regime. The stationary property of the deformations for both the atomic and the nuclear emissions suggests that the observed anomaly is attributed to the same origin of the collective $^{103m}$Rh nuclei in interaction.

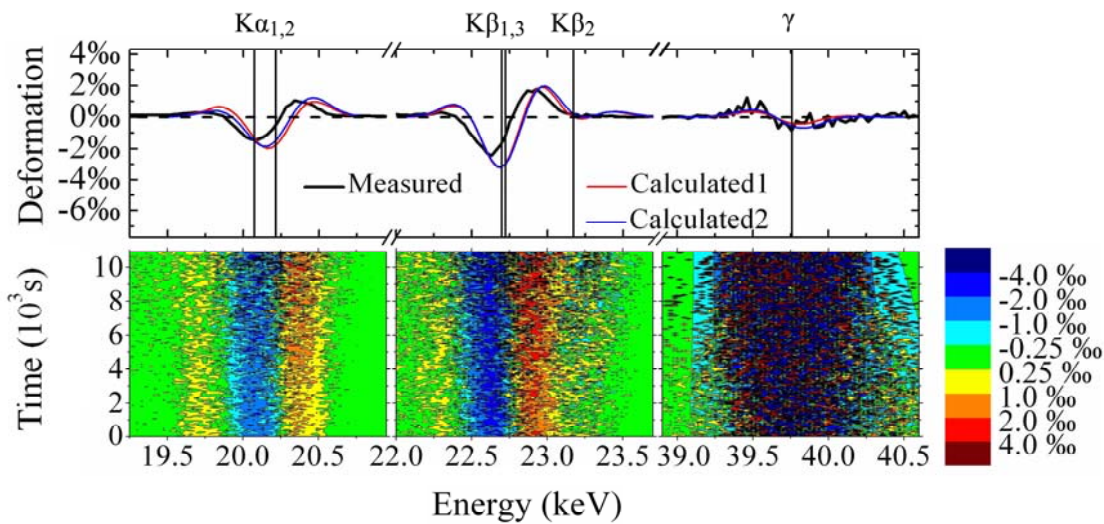

Figure 5. (Color online) Spectral deformations (top) and the corresponding time evolution behavior (bottom) for the three bands of Kα, Kβ, and γ analyzed by (1) and (2). The data are taken in the linear regime by 210 Hz without the Cu filter. In the time-resolved plot, the data for the analysis of spectral deformations are in counts per minute within the time span of measurement in three hours. The amplitude of deformation is presented by the color shown on the right hand side. In the top figure, the two calculated curves (1: red curve for D model and 2: blue curve for T model) with parameters listed in Table 3 is compared with the experimental data (black curve) accumulated in three hours. The two vertical lines at the position of Kα are for Kα$_1$ and Kα$_2$. At the position of Kβ, the two almost coincided vertical lines are for Kβ$_1$ and Kβ$_3$, and the third one is for Kβ$_2$. The location of γ at 39.76 keV is also indicated by the vertical line.

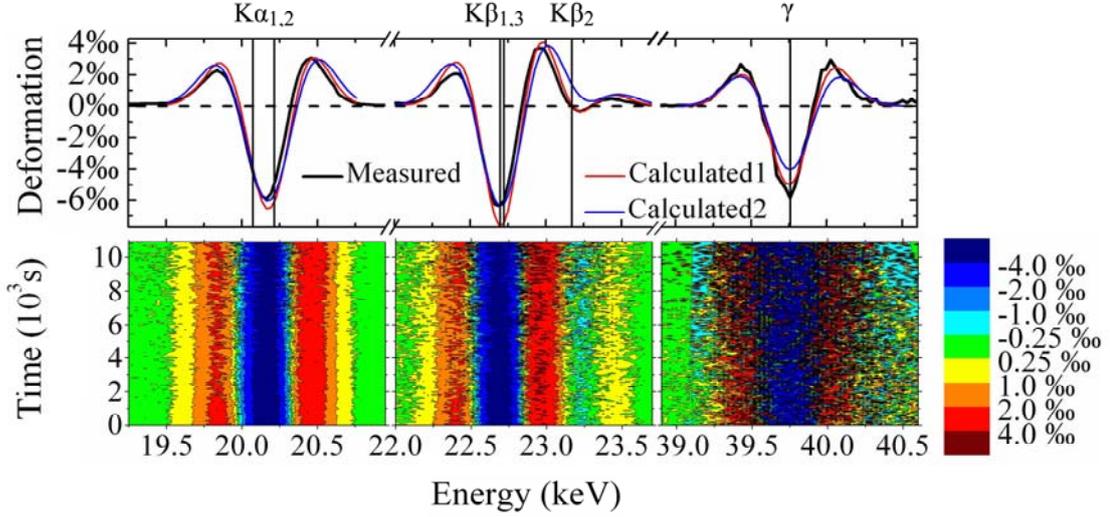

Figure 6. (Color online) Spectral deformations (top) and the corresponding time evolution behavior (bottom) for the three bands of Kα, Kβ, and γ. The data are taken by 295 Hz in the nonlinear regime without the Cu filter. The convention of this figure is the same as that described in figure 5.

**Table 3.** Splitting energies, $\Delta E$, for Kα, Kβ and γ peaks obtained by the analysis according to (2). Two of the measurements by 210 and 295 Hz as shown in figures 5 and 6 are without the Cu filter, whereas the other two measurements by 105 and 260 Hz as discussed in figure 4 are with the Cu filter, for which the effect of Cu absorption is accounted for in the model. The D model is for $A_{c,i} = 0$, while the T model is assuming $A_{c,i} = 80\%$, as discussed in text.

| $\Delta E$ (eV) | Kα | | Kβ | | γ | |
|---|---|---|---|---|---|---|
| Model | D | T | D | T | D | T |
| 105 Hz | 38 | 82 | 44 | 111 | 17 | 46 |
| 210 Hz | 44 | 100 | 55 | 136 | 21 | 50 |
| 260 Hz | 40 | 95 | 48 | 120 | 11 | 59 |
| 295 Hz | 85 | 212 | 92 | 235 | 75 | 165 |

## 4. Conclusion

The $^{103m}$Rh emission spectra from the polycrystalline sample excited by bremsstrahlung irradiation exhibit several interesting properties. The inversion density of $^{103m}$Rh beyond $3\times10^{11}$ cm$^{-3}$ is demonstrated to increase nonlinearly with the bremsstrahlung exposure. Stationary triplet splittings of Kα, Kβ and γ with the same magnitude of splitting energy for these characteristic emissions are observed. The γ splitting energies remain roughly at the same magnitude in the linear regime, but are significantly enhanced for the measurement in the nonlinear regime. The asymmetrical part of the K splitting may be attributed to the satellite lines created by the multiple ionizations. However, the γ splitting of $^{103m}$Rh is never observed by the similar multipolar transition of $^{109m}$Ag. This will be addressed in our future works. More ongoing measurements have been or will be performed to investigate the interesting properties of the anomalous characteristic emissions of $^{103m}$Rh excited by bremsstrahlung irradiation with different experimental conditions.

**Acknowledgments**

We give special thanks to the accelerator team of department of engineering physics, Tsinghua University. This work is supported by the NSFC grant 10675068.

**References**
[1] Cheng Y, Xia B, Liu Y-N, Jin Q-X 2005 *Chin. Phys. Lett.* **22** 2530
[2] Cheng Y, Xia B, Tang C-X, Liu Y-N, Jin Q-X 2006 *Hyperfine Interactions* **167** 833
[3] Hubbell J H and Seltzer S M, 1995 *Tables of X-ray mass attenuation coefficients and mass energy-absorption coefficients 1keV to 20MeV for elements 1-92 and 48 additional substances of dosimetric interest,* (National Institute of Standards and Technology Internal Report, NISTIR 5632)
[4] Berger M J and Hubbell J H, Seltzer S M, Chang J, Coursey J S, Sukumar R, and Zucke D S 1998 *XCOM: Photon Cross Sections Database* http://www.physics.nist.gov/PhysRefData/Xcom/Text/XCOM.html
[5] Firestone R B (ed.) 1999 *Table of Isotopes*, 8$^{th}$ Eds. (John Wiley & Sons, New York)
[6] van Eijk C W E and Wijnhorst J 1977 *Phys. Rev. C* **15** 1068
[7] van Eijk C W E, Wijnhorst J and Popelier M A 1979 *Phys. Rev. C* **19** 1047

# Supporting online material

Observations on spectral deformations of $^{103m}$Rh excited by bremsstrahlung

**Y Cheng[1], B Xia[1], Q-X Jin[1], and C P Chen[2]**

[1)]Department of engineering Physics, Tsinghua University, 100084, Beijing, China
[2)]Department of Physics, Peking University, 100871, Beijing, China

E-mail: yao@tsinghua.edu.cn

In order to show that the measurement by the bremsstrahlung irradiation with repetition frequency of 295 Hz presented in the submitted paper is not an isolated example in the nonlinear regime, a brief account is given to provide more experimental evidences of measurements in the nonlinear regime. Since these measurements are carried out by using another low-energy high purity germanium (HPGe) detector and more interesting properties are revealed by these measurements than those in the linear regime, we choose to put forward only the relevant and supportive results in this supporting material. A complete and more detailed report is under preparation and will be submitted in a separate article.

Five figures are provided to show, (a) in figure S1, the calibration on the detection efficiency for the two different HPGe detectors by CANBERRA, of models GL0210P with an active detector area of 200 mm$^2$ (denoted as detector A) and GL0510P with an active detector area of 500 mm$^2$ (denoted as detector B), (b) in figure S2, the K$\alpha$ luminosity of $^{103m}$Rh versus the exposure rate of bremsstrahlung irradiation, (c) in figure S3, the splitting energy versus the K$\alpha$ luminosity (inversion density) of $^{103m}$Rh for the K$\alpha$, K$\beta$, and $\gamma$ peaks at room temperature, (d) in figure S4, the resolvable triplet splitting in the highly nonlinear regime, and (e) in figure S5, the triplet splitting with repetition frequency of 295 Hz presented in the submitted paper is not resolvable.

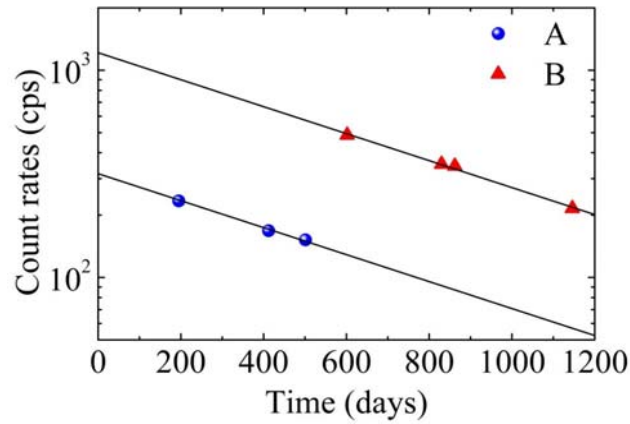

Figure S1. (Color on line) Calibration of the detection efficiency between the two detectors used in the paper (red filled triangle, CANBERRA GL0510P, detector B) and in this supporting material (blue filled circle, CANBERRA GL0210P, detector A). The radioactive source is $^{109}$Cd with a half life of 462.6 days. The count rate is for the K$\alpha$ peak at 22 keV, which are in the energy range of interest for the study on the emissions of $^{103m}$Rh. The conversion ratio is obtained from the two calibration lines as 3.834±0.001.

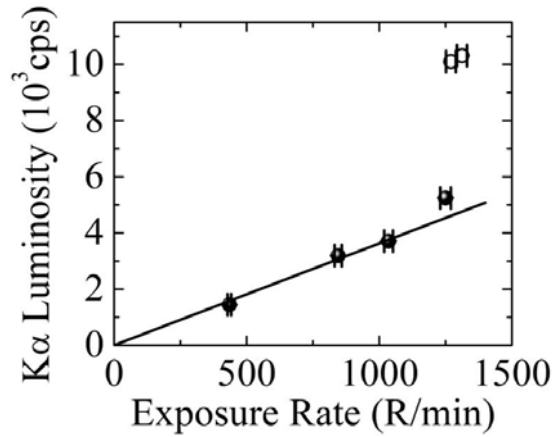

Figure S2. K$\alpha$ luminosity of $^{103m}$Rh at the beginning of measurement in variation with the recorded value of the exposure rate in Röntgen per minute (R/min). The filled circle is for the data taken at room temperature by the HPGe detector B (CANBERRA GL0510P) in the present experiment. The repetition frequencies of the electron beam to generate the specified bremsstrahlung exposure rate are 105, 210, 260, and 295 Hz. The open circles, showing highly nonlinear dependence on the exposure rate, are taken at room temperature by the HPGe detector A (CANBERRA GL0210P). The solid line is for the linear fit from the three data points below 1200 R/min.

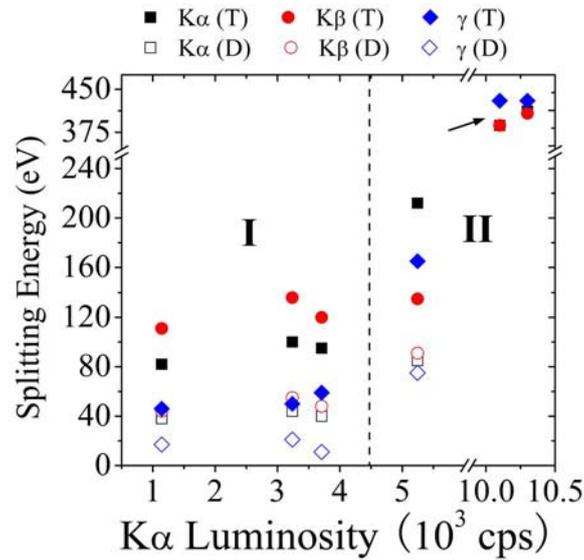

Figure S3. (Color on line) Splitting energy versus the Kα luminosity of $^{103m}$Rh analyzed for the three peaks of Kα (black filled squares with triplet model denoted by T, black open squares with doublet model denoted by D), Kβ (red filled circles with T, red open circles with D), and γ (blue filled diamonds with T, blue open diamond with D). The data points correspond to the six measurements presented in figure S2. For the first four data points in the left, we show the analysis results by models T and D, which are discussed in the paper. Two data points with Kα luminosity > $10^3$ counts per second (cps) exhibit the splitting energy > 400 eV, of which the T model is fully resolved by the detector. Therefore, the analysis with the D model is not necessary for these two data points. The three points on the left hand side are in the linear regime I and the three data points on the right are in the nonlinear regime II. The first four data points on the left hand side taken by the HPGe detector B (CANBERRA GL0510P), are presented in the paper, while the two data points on the right hand side are taken by the HPGe detector A (CANBERRA GL0210P), and will be reported in the future with detailed analysis. This figure reveals that the energy splitting for the characteristic emissions is not detector dependent.

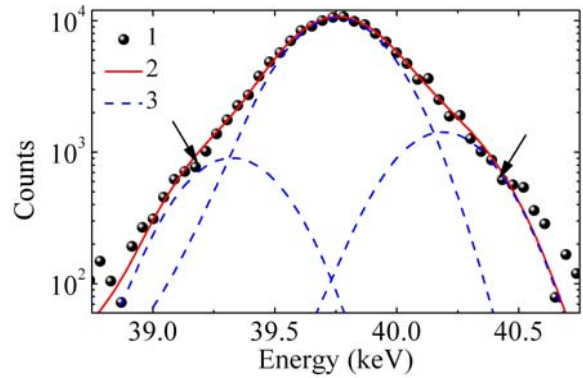

Figure S4. (Color on line) Triplet splitting of the γ peak at 39.76 keV for the data point with the Kα luminosity of $10.3\times10^3$ cps in figure S3, Black filled circles (1) are the measured counts of experimental data (43.5 eV per channel). Solid line (2) in red color is the fitting result by the triplet splitting model, by which the three peaks are with the normal profile, as shown by the dashed lines (3) in blue color. Besides the broadening of the peak, the triplet feature is clearly demonstrated by two kinks indicated by the arrows at the level of 800 counts. The Kα-Kα pile-up (40.4 keV) at the right shoulder of γ peak is removed by the off-line data analysis.

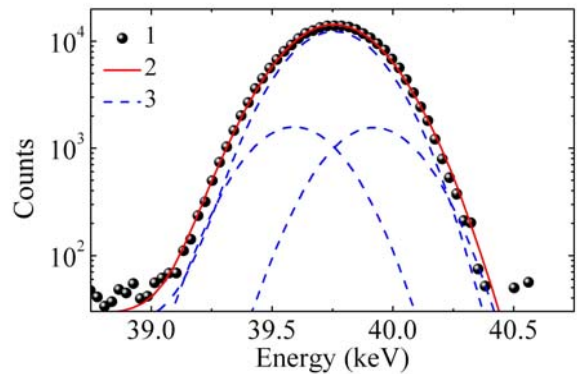

Figure S5. (Color on line) Broadening of the γ peak at 39.76 keV for the data point with the Kα luminosity of $5.2\times10^3$ cps in figure S3, It is fitted by the triplet splitting model. Black filled circles (1) are the measured counts of experimental data (29.4 eV per channel). Solid line (2) in red color is the fitting result by the triplet model. The three peaks are assumed with the distribution of normal profile, as shown by the dashed lines (3) in blue color. There is no clear indication of triplet feature revealed by the original data except for the broadening. In this case, the splitting energy by the T model is 165 eV, much smaller than the FWHM of the central profile. The Kα-Kα pile-up (40.4 keV) at the right shoulder of γ peak is removed by the off-line data analysis.